\documentclass[10pt,conference]{IEEEtran}

\ifCLASSINFOpdf

\else

\fi

\usepackage{array}

\usepackage{pst-node}
\usepackage{lipsum}
\usepackage{epstopdf}
\usepackage{graphicx}
\usepackage{amsmath}
\usepackage{amssymb}
\usepackage{xcolor}
\usepackage{xfrac}
\usepackage{braket}
\usepackage{subcaption}
\usepackage{comment}
\usepackage{float}
\usepackage{mathtools}
\usepackage{tikz}
\usetikzlibrary{quantikz2}
\usepackage[]{mdframed}
\usepackage{multirow}

\newcommand{\abs}[1]{\lvert #1 \rvert}

\newcommand{\kett}[1]{| #1\rangle}

\newcommand{\expec}[3]{\langle #1\lvert #2 \rvert #3\rangle}
\DeclarePairedDelimiter{\ceil}{\lceil}{\rceil}

\begin{document}
\title{mRNA secondary structure prediction using utility-scale quantum computers}

\author{\IEEEauthorblockN{Dimitris Alevras\IEEEauthorrefmark{1},
Mihir Metkar\IEEEauthorrefmark{3},
Takahiro Yamamoto\IEEEauthorrefmark{2},
Vaibhaw Kumar\IEEEauthorrefmark{1}, 
Triet Friedhoff\IEEEauthorrefmark{1},\\
Jae-Eun Park\IEEEauthorrefmark{1},
Mitsuharu Takeori\IEEEauthorrefmark{2},
Mariana LaDue\IEEEauthorrefmark{1},
Wade Davis\IEEEauthorrefmark{3}, and
Alexey Galda\IEEEauthorrefmark{3}}
\IEEEauthorblockA{\IEEEauthorrefmark{1}IBM Quantum, New York, USA}
\IEEEauthorblockA{\IEEEauthorrefmark{2}IBM Quantum, Tokyo, Japan}
\IEEEauthorblockA{\IEEEauthorrefmark{3}Moderna, Cambridge, USA}
}

\bstctlcite{IEEEexample:BSTcontrol}

\maketitle

\begin{abstract}
Recent advancements in quantum computing have opened new avenues for tackling long-standing complex combinatorial optimization problems that are intractable for classical computers. Predicting secondary structure of mRNA is one such notoriously difficult problem that can benefit from the ever-increasing maturity of quantum computing technology. Accurate prediction of mRNA secondary structure is critical in designing RNA-based therapeutics as it dictates various steps of an mRNA life cycle, including transcription, translation, and decay. The current generation of quantum computers have reached utility-scale, allowing us to explore relatively large problem sizes. In this paper, we examine the feasibility of solving mRNA secondary structures on a quantum computer with sequence length up to 60 nucleotides representing problems in the qubit range of 10 to 80. We use Conditional Value at Risk (CVaR)-based VQE algorithm to solve the optimization problems, originating from the mRNA structure prediction problem, on the IBM Eagle and Heron quantum processors. To our encouragement, even with ``minimal'' error mitigation and fixed-depth circuits, our hardware runs yield accurate predictions of minimum free energy (MFE) structures that match the results of the classical solver CPLEX. Our results provide sufficient evidence for the viability of solving mRNA structure prediction problems on a quantum computer and motivate continued research in this direction. 

\end{abstract}

\IEEEpeerreviewmaketitle

\section{Introduction}\label{sec:introduction}

Messenger RNA (mRNA) plays a pivotal role in the central dogma of molecular biology, acting as the intermediary between the genetic blueprint in DNA and the functional molecule, protein. The functionality of mRNA is inherently linked to its secondary structure, which arises from intramolecular base pairing and folding patterns~\cite{vicens2022thoughts}. Understanding these structures is crucial for elucidating the mechanisms of gene regulation, translation, and degradation kinetics, and inter-biomolecular interactions~\cite{metkar2024tailor}. Additionally, the emergence of mRNA-based therapeutics has ignited a paradigm shift in the landscape of medicine, offering novel solutions for the treatment of various diseases~\cite{qin2022mrna}. This transition from understanding mRNA's fundamental biological role to harnessing its therapeutic potential marks a significant milestone in biomedical research and holds immense promise for revolutionizing healthcare practices. However, predicting mRNA secondary structures is a formidable combinatorial problem due to the vast number of possible configurations and their associated energy states~\cite{mathews2010folding}. Solving this problem accurately and efficiently is essential for advancing our knowledge in genetics, aiding in the design of mRNA-based drugs, and developing gene therapies.

The problem of RNA secondary structure prediction with pseudoknots is known to be NP-complete~\cite{lyngso2000rna}. State-of-the-art classical approaches for the secondary structure prediction of mRNA have primarily relied on dynamic programming algorithms, such as the Zuker algorithm~\cite{zuker198920} and its widely used implementations, MFold~\cite{zuker2003mfold} and ViennaRNA~\cite{schuster1997rna}, to name a few. These methods calculate the minimum free energy (MFE) structures by systematically evaluating all possible base pairings and applying thermodynamic models to estimate their stability~\cite{zuker1989finding}. Subsequent enhancements have incorporated stochastic context-free grammars (SCFGs), as seen in tools like Infernal~\cite{nawrocki2013infernal}, which align sequences to RNA structure profiles to predict conserved structures across related sequences. The integration of machine learning techniques was implemented in algorithms like ContextFold~\cite{zakov2011rich} that utilize both local and global contextual features to improve prediction accuracy. Despite these advancements, classical approaches are still constrained by the combinatorial explosion of possible structures, lack of generalization to novel RNAs~\cite{szikszai2022deep}, and the approximations required to make the problem tractable~\cite{zhang2020linearpartition}. As a result, there is a growing interest in leveraging quantum computing to overcome these limitations and achieve more accurate predictions of mRNA secondary structures.

Classical hardness of mRNA secondary structure prediction naturally positions it as a compelling application for quantum computers. Despite still being relatively scarce, recent developments have shown promising advancements in this area. In Refs.~\cite{fox2022rna, zaborniak2022qubo}, quantum annealing-based hardware was used to minimize the energy function that corresponds to the folding energy landscape of mRNA molecules. However, quantum annealers are not universal quantum processors, which limits the set of algorithms amenable to this technology compared to the universal gate-based quantum computers. In another recent paper~\cite{jiang2023predicting}, the Quantum Approximate Optimization Algorithm (QAOA)~\cite{farhi2014quantum} was used to solve the secondary structure prediction problem for toy-model RNA sequences using 12 qubits on a simulator and 4 qubits on a gate-based quantum processor. 

Algorithms such as QAOA and Variational Quantum Eigensolver (VQE) fall under the class of variational quantum algorithms that can be implemented on the current generation of gate-based quantum computers. Guided by a classical optimization feedback loop, the goal of such algorithms is to find the quantum state that exhibits sufficient overlap with the ground state of the optimization problem. Success of such schemes depends heavily on the choice of a suitable ansatz, objective function, and efficient classical optimizer, as highlighted in the literature. However, most of such studies are either based on numerical simulations or examine relatively small problems on quantum hardware, with only a few notable exceptions~\cite{harrigan2021quantum, he2023alignment, barron2023provable, sack2024large, miessen2024benchmarking}. 

As universal quantum processors surpassed one hundred qubits, we have entered the ``utility'' era of quantum computing~\cite{kim2023evidence}, where quantum computers can perform reliable computations on problem sizes beyond the range that can be handled by brute-force classical schemes. Encouraged by these developments, in this work we examine the feasibility of solving relatively longer mRNA sequences on a universal utility-scale quantum processor. The goal here is to demonstrate that variational quantum algorithms can be fully and successfully executed on such hardware, with hundreds of quantum circuit calls during the hyperparameter training step along with basic error mitigation, yielding accurate results that match those obtained with classical optimization schemes.

In this work, we use the Conditional Value at Risk Variational Quantum Eigensolver (CVaR-VQE), proposed as a modification to the traditional VQE algorithm, that employs CVaR values of the energy eigenfunctions as the objective function, to achieve better convergence~\cite{barkoutsos2020improving}. Additionally, we choose Nakanishi-Fujii-Todo (NFT)~\cite{nakanishi2020sequential} as a classical optimizer along with the hardware-efficient ``two-local'' ansatz. The combination of the NFT optimizer and two-local ansatz has been shown to be relatively robust against hardware noise and exhibits better convergence properties~\cite{oliv2022evaluating}.  

First, we formulate the secondary structure prediction problem as a binary optimization problem. We then use CPLEX~\cite{cplex} to solve the formulated problem. We provide numerical evidence that as the problem size increases, classical schemes like CPLEX show an extremely unfavorable runtime scaling. Next, we run these problems using both quantum hardware and simulations, and present results on RNA sequence lengths ranging from 15 to 42 bases representing optimization problems in the range of 10 to 80 qubits. To the best of our knowledge, our study is the first to examine RNA sequences up to 42 nucleotides on a quantum processor.

The paper is organized as follows. In Section~\ref{sec:problem}, we offer a formal definition of the problem and give a brief description of the structures that provide more or less stability. In Section~\ref{sec:classical}, we briefly describe the current classical methods and describe state-of-the-art classical algorithms, with focus on  dynamic and mathematical programming approaches. We also provide our formulation of the problem as a mathematical programming problem. Section~\ref{sec:quantum} focuses on the quantum formulation of the problem and describes the CVaR-VQE method used in this work. Section~\ref{sec:results} contains the results of our experiments, using both classical simulators and quantum processors. Finally, in Section~\ref{sec:conclusions}, we conclude with a summary of our results and an outlook.

\begin{figure}[!ht]
\centering
\includegraphics[width=.5\textwidth]{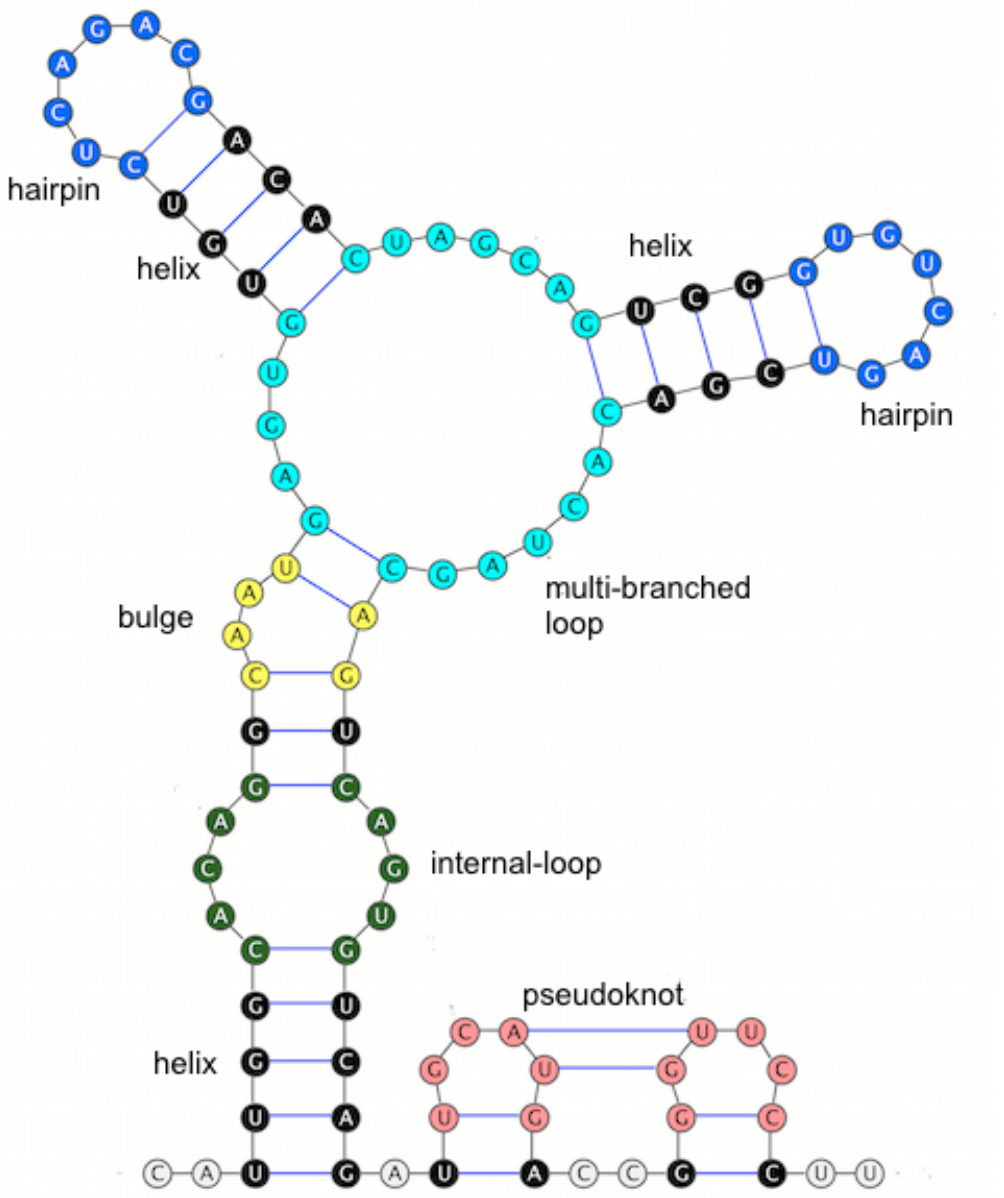}
\caption{An RNA secondary structure illustrating five kinds of secondary structural elements: hairpin loops (blue), bulge (yellow), helix (black), internal loop (green) and multi-branched loop (cyan) and a pseudoknot (pink)~\cite{mamuye}.}
\label{fig:mrna_graph_grammar}
\end{figure}

\section{Problem definition}\label{sec:problem}

\begin{figure}[!ht]
\centering
\includegraphics[width=.5\textwidth]{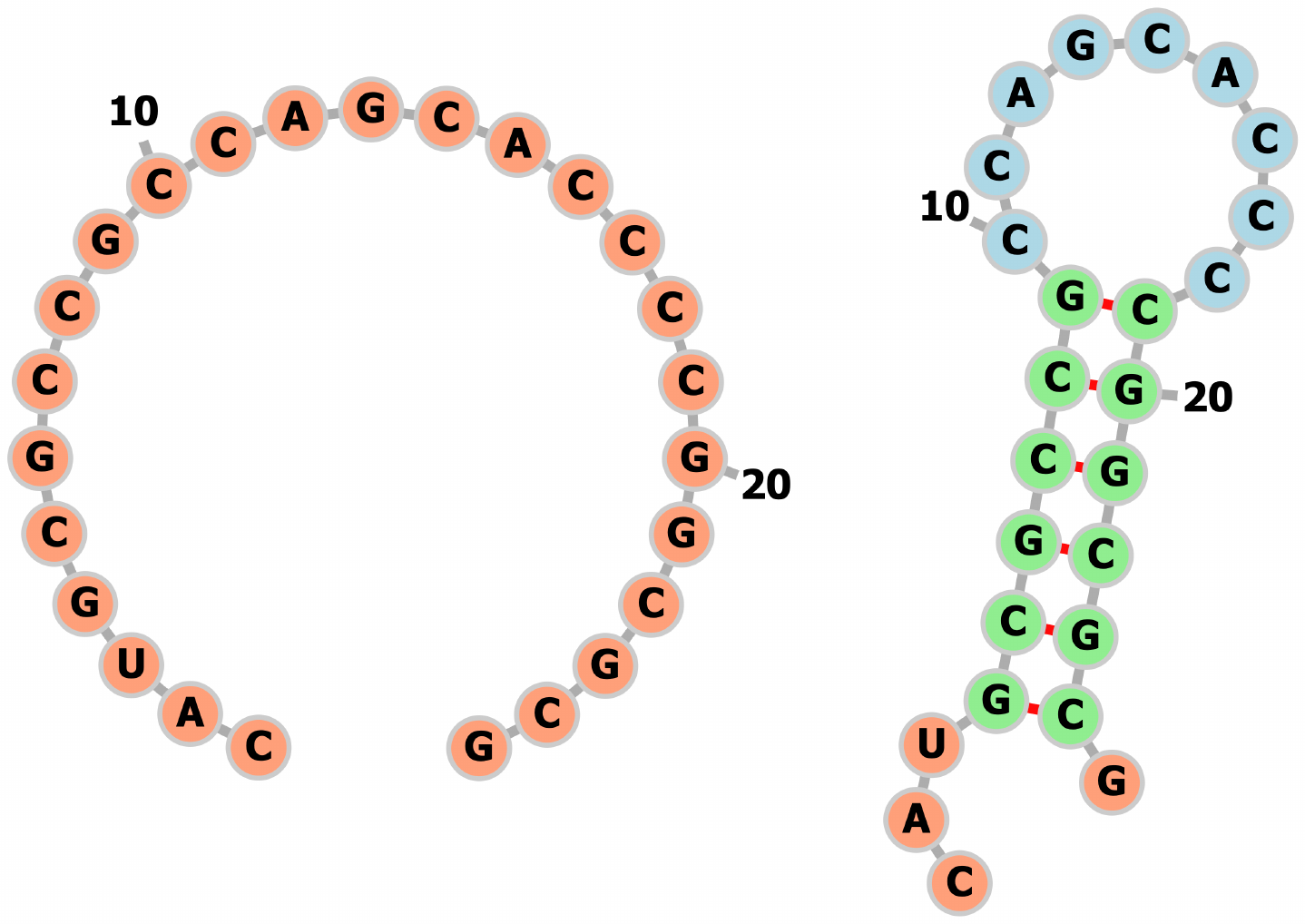}
\caption{Example of an RNA sequence and its predicted MFE structure demonstrating its folded state. Optimal folding calculated by ViennaRNA RNAfold server. Graphic generated with ViennaRNA forna server~\cite{vienna}.}
\label{fig:mrna}
\end{figure}

RNA commonly is a polymer of four nucleotides (bases): adenine (A), cytosine (C), guanine(G), and uracil (U). Under physiological conditions, RNA adopts compact structures driven by hydrophobic, base-stacking, and base-pairing interactions~\cite{vicens2022thoughts}. Typically these interactions facilitate the formation of an A-form helix containing canonical Watson-Crick-Franklin base-pairs. The folded RNA conformation can be depicted as secondary structures, comprising various secondary structural elements that arise from base-pairing and base-stacking interactions (see Fig.~\ref{fig:mrna_graph_grammar}). Understanding RNA secondary structure is crucial for gaining insights into its function and designing RNA-based therapeutics~\cite{metkar2024tailor}.
The RNA secondary structure prediction problem is the problem of finding the most stable folding of the sequence of bases~\cite{durbin1998biological}. The exact characteristics of this ``optimal'' folding are not completely understood, making approximate models well-suited for this purpose. In fact, in certain cases, there may exist multiple optimal structures, underscoring the value of methods that can produce numerous high-quality solutions.

According to thermodynamic principles, formation of stem-loop structure is feasible if the entropic costs of closing loops and ordering nucleotides in a helix are offset by free energy released due to base-stacking and pairing interactions~\cite{andronescu2007efficient}. As a result, longer helices, or stems, are typically more stabilizing for secondary structures, whereas loops and single-stranded regions tend to introduce instability. To predict the most favorable folding, a computational model must determine the structure with the lowest free energy (MFE) compared to the unfolded state, which is the conformation that RNA is most likely to adopt under physiological conditions~\cite{andronescu2007efficient}. 

\section{Classical approaches}\label{sec:classical}

The most common way of predicting secondary structure is to minimize the free energy change upon moving from an unfolded state to the most thermodynamically stable folded state, see Fig.~\ref{fig:mrna}~\cite{zuber2022nearest}. This secondary structure which has the lowest possible energy is defined as the MFE structure. Because MFE reflects the most thermodynamically stable structure, many approaches for predicting secondary structure concentrate on identifying structures through MFE minimization. However, this problem, viewed as an optimization problem, is generally NP-hard. There are approximation schemes designed to optimize simplified models by excluding specific structures like pseudoknots (see Fig.~\ref{fig:mrna_graph_grammar}) and focusing only on a subset of possible structures~\cite{durbin1998biological, qfold}. These include dynamic programming algorithms (described below), along with genetic algorithms~\cite{chen2000} and machine learning algorithms~\cite{zhao2021}, to name a few. See Ref.~\cite{Seetin2012} for a review of classical methods for RNA structure prediction.

The potential of quantum computing to tackle this problem stems from two key aspects:
\begin{itemize}
    \item A Quadratic Unconstrained Binary Optimization (QUBO) problem can be converted to the problem of finding the lowest energy eigenstate of an Ising Hamiltonian and thus can be solved by a quantum algorithm;
    \item The secondary structure prediction problem can be formulated as a QUBO.
\end{itemize}
Previous efforts have focused on developing an objective function to minimize over binary variables, with subsequent optimization carried out on a quantum annealing hardware~\cite{fox2022rna,zaborniak2022qubo}.
In this work, we use mathematical programming to represent the problem in binary variables and then convert the resulting program to QUBO, which we solve on a universal quantum computer.

\subsection{Dynamic programming algorithms}

One approach to approximate the ``optimal'' folding is to maximize the number of paired bases. This can be accomplished by the dynamic programming algorithm of Nussinov~\cite{durbin1998biological}. The algorithm is based on the fact that the maximum number of paired bases for sequence $i,\ldots,j$ can be calculated from the maximum number of pairs for smaller length sequences. There are only four possible ways to construct nested base-pairs from smaller sub-sequences~\cite{Eddy2004}. It is therefore possible to recursively calculate the maximum number of pairs. Even though maximizing the number of paired bases is not sufficient, Nussinov's algorithm provides the basic mechanism of subsequent dynamic programming algorithms, such as Zuker's algorithm~\cite{zuker1981optimal}; one has to replace the number of maximum pairs in Nussinov's algorithm with a scoring function (such as MFE) and calculate the score for each sub-sequence. The best approximation to what ``optimal'' means is the one that results in the best score (e.g., MFE) for the sequence. 

Dynamic programming algorithms do not account for all structures, especially a specific type of structure known as a pseudoknot, shown in Fig.~\ref{fig:mrna_graph_grammar}. While not present in all mRNA sequences, and particularly relatively short sequences considered in this work, pseudoknots are crucial for the function of several important RNA elements, including regulation of translation and splicing~\cite{draper2000rna} and the binding of small molecules~\cite{gilbert2008structure, klein2009cocrystal, spitale2009structural}. However, for many practical purposes, approximation methods that do not explicitly account for pseudoknots can still provide reasonable and efficient predictions of mRNA secondary structures. The current classical algorithms utilized in software achieve 65-73\% accuracy in predicting secondary structures observed in the laboratory~\cite{mathews1999}.

\subsection{Mathematical programming models}

Among classical schemes, mathematical programming models and integer linear programming, in particular, are often used to solve the RNA folding problem~\cite{gusfield2019integer}. Although these models end up being very large in the number of variables and constraints, the classical solvers available today are sophisticated and powerful enough to handle problems of reasonable size. A significant benefit of integer linear programming lies in its flexibility to accommodate modifications to the problem, such as incorporating new structural motifs like pseudoknots, without disrupting the method, unlike dynamic programming algorithms.

Several ways have been proposed to represent the secondary structure prediction problem, defining a variable to be:
\begin{itemize}
    \item  a pair between bases $i$ and $j$  represented as $(i,j)$
    \item  a stack of pairs  $(i,j),(i+1,j-1),\ldots,(i+k,j-k)$ represented as $(i,j,k)$
    \item  a stack of two consecutive pairs, also known as a quartet, represented as $(i,j,i+1,j-1)$.
\end{itemize}

Using a pair between $i$ and $j$ as a variable works well in solving the maximum pair problem (like Nussinov's algorithm) but makes it difficult to model more complex requirements and constraints.
In terms of scaling, the number of variables of the problem with respect to the sequence length scales best when using quartets and worse when using pairs.

We use mathematical programming to model the problem as a binary optimization problem. Each of the positions in an RNA sequence $1,\ldots,n$ consists of one of the bases $U,A,C,G$. A pair between two bases in positions $i$ and $j$ in the sequence is allowed if the corresponding bases form a valid pair. The valid pairs are: $\{(AU), (UA), (CG),(GC),(GU),(UG)\}$. The variables are defined for each quartet~\cite{gusfield2019integer} which is two consecutive pairs (also referred to as stacked pairs):
$$x(i,j,i+1,j-1)=\left \{ 
\begin{array}{rl}
1 & \text{if $(i,j)$, $(i+1,j-1)$ are made}\\
0 & \text{otherwise}
\end{array} 
\right .
$$ 

These variables are generated if the pairs are valid. The following data pre-processing steps are performed to prepare the data for the formulation:
\begin{itemize}
    \item check if two pairs can create a quartet (can nest)
    \item check two pairs for having the same base
    \item checks two pairs for crossing
    \item create a list of quartets (variables of the problem)
    \item check two quartets for crossing
\end{itemize}

Two quartets cannot be both selected if there are crossing pairs between them. Also, each base can be paired only with one other base. These conditions result in constraints of the form:
$$x(i,j,i+1,j-1) + x(k,\ell,k+1,\ell-1) \le 1\,.$$
Since the variables are binary, the constraint above is equivalent to requiring that:
$$x(i,j,i+1,j-1)  x(k,\ell,k+1,\ell-1) = 0\,,$$
which can be included in the objective function as a product of the two variables with a penalty, thus preventing both of them to be 1. We note that depending on the algorithm used to solve the problem, we may want to keep one (with constraints) or the other (without constraints) formulation. In our experiments, we utilize CPLEX, which demonstrates superior performance when explicit constraints are included in the formulation.

The objective function has the following components:
\begin{itemize}
    \item Each variable (quartet) has a free energy as given in Ref.~\cite{turner-mathews}.
    \item Consecutive quartets are preferred since they provide more stability. We model this, by adding the product of the corresponding variables in the objective function with a reward.
    \item Certain structures are discouraged. In this work we consider the $(UA)$ pair ending penalty as given in Ref.~\cite{turner-mathews}. We implement this by adding  variables that form such ending pair to the objective function with a penalty.
\end{itemize}

For brevity of notation, we use $q_i$ to denote the $i^{th}$ quartet variable. Let $Q$ be the set of valid quartets,  $QC$ the set of all pairwise crossing quartets, $QS(q_i)$ the set of quartets that can be stacked with quartet $q_i$, and $\mathit{QUA}$ the set of stacked quartets that have a $(UA)$ end pair. The sets $Q$, $QC$, $QS(q_i)$ and $\mathit{QUA}$, are determined based on the given sequence in a pre-processing step. Each quartet $q_i \in Q$ is formed by two consecutive (stacked) pairs $p^i_1$ and $p^i_2$. The MFE  $e_{q_i}$ for quartet $q_i$ is the MFE of $p_1$ followed by $p_2$~\cite{turner-mathews}. Two quartets $q_1, q_2 \in Q$ that are stacked are rewarded by a reward $r$. If a stack ends in a $(U,A)$ pair there is a penalty $p$. The objective function is thus:
\begin{equation} \label{eq:obj}
    \min \sum_{q_i \in Q} e_{q_i} q_i  +
r\sum_{q_i \in Q}{\sum_{q_j \in QS(q_i)}}q_i q_j +
p \sum_{q_i \in Q} \sum_{q_j \in \mathit{QUA}} q_i (1-q_j)\,.
\end{equation}

The constraints of the problem implement the requirement that  if two quartets $q_1, q_2 \in Q$ are crossing (i.e., contain crossing pairs) they cannot be both selected:
\begin{equation}\label{eq:constraint}
    q_i + q_j \le 1\,, \quad \quad \forall q_i, q_j \in QC\,.
\end{equation}

\begin{figure}[!ht]
\centering
\includegraphics[width=.5\textwidth]{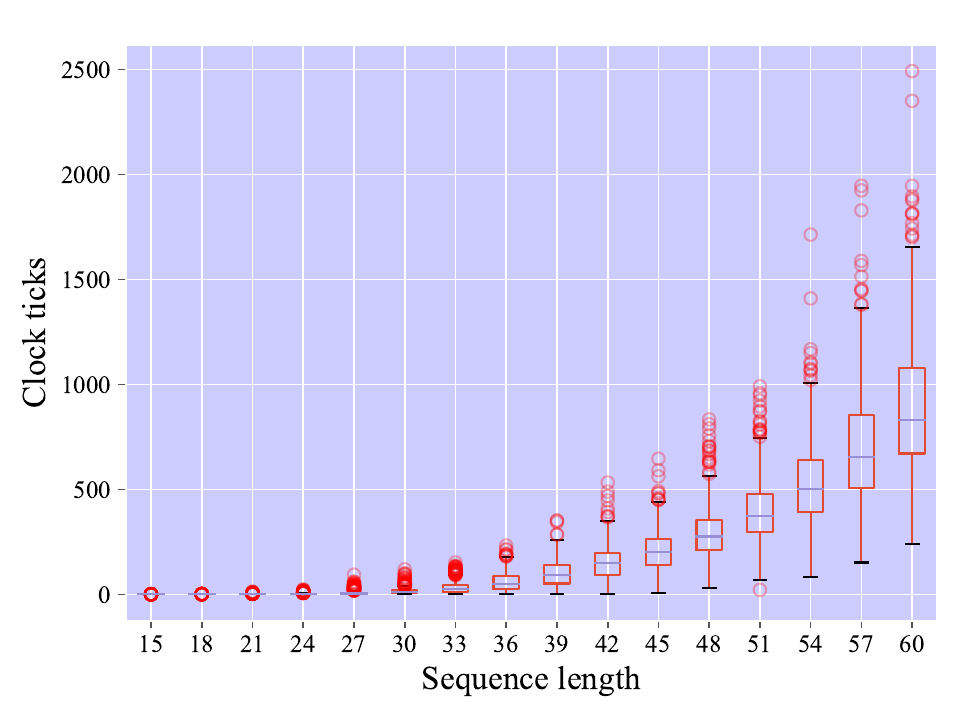}
\caption{Time scaling (internal clock ticks) for CPLEX as a function of sequence length for 10,000 random RNA sequences.}
\label{fig:cplex_scale}
\end{figure}

The resulting problem is a quadratic binary optimization problem. We use CPLEX as a classical solver to solve this problem and create a baseline. As noted above, for larger sequences the number of constraints in this formulation is quite large. However, modern classical solvers, such as CPLEX, can deal with the size of the problem. In our experiments, we generated 10,000 sequences of length varying between 15 and 60 nucleotides,  which  resulted in quadratic binary problems with up to 327 variables and up to 38,946 constraints. CPLEX was able to find  optimal solutions to all instances within a few seconds per instance. In Fig.~\ref{fig:cplex_scale} we plot the time CPLEX takes to solve a problem, measured in internal clock ticks, as a function of length of RNA sequences. Our dataset contains over 600 problem instances per sequence length
The box plot suggests that the time to solve the hardest instances grows exponentially with sequence length. As a result, in the mathematical programming formulation of the problem, classical solvers will struggle to find a solution within a reasonable time frame even for sequences of a few hundred nucleotides.

\section{Quantum approach}\label{sec:quantum}

\subsection{QUBO formulation}\label{sec:QUBO_modeling}

The mathematical programming formulation described above is a quadratic constrained binary optimization problem. The corresponding problem Hamiltonian can be obtained by transforming the constrained problem to a QUBO problem. Other approaches~\cite{zaborniak2022qubo, fox2022rna} formulate the QUBO directly and train the model to find the appropriate penalties and rewards for the objective function. Here we use the Qiskit~\cite{Qiskit} available methods to convert the problem to a QUBO and calculate the penalties for the constraints. 

As previously mentioned, the constraints in Eq.~\ref{eq:constraint} can be incorporated into the objective function (Eq.~\ref{eq:obj}) through the introduction of a slack variable and a penalty term. However, due to the form of the constraints and the binary nature of the variables, the introduction of additional slack variables is unnecessary. After adding a penalty $t$ for the constraints, the final objective function of the resulting QUBO takes the following form:
\begin{equation} \label{eq:qubo}
\begin{array}{lcl}
    \min & \ & \displaystyle \sum_{q_i \in Q} e_{q_i} q_i   \\
& \ &+ \displaystyle r\sum_{q_i \in Q}{\sum_{q_j \in QS(q_i)}}q_i q_j \\
& \ &+ \displaystyle p \sum_{q_i \in Q} \sum_{q_j \in \mathit{QUA}} q_i (1-q_j) \\
& \ &+\displaystyle t \sum_{q_i,q_j \in QC} q_i q_j\,, \\
\text{s.t.} & \ & q_i \in \{0,1\} \ \forall q_i \in Q\,.

\end{array}
\end{equation}

\subsection{CVaR-based VQE}
\label{cvar-vqe}
In this work, we use Conditional Value at Risk (CVaR)-based VQE~\cite{barkoutsos2020improving, robert2021resource} to solve the QUBO problems arising from the formulation discussed in Section~\ref{sec:QUBO_modeling}. VQE is a hybrid quantum-classical optimization algorithm which uses a unitary $U(\boldsymbol{\theta})$, characterized by the rotation angles $\boldsymbol{\theta}$, to prepare a parameterized quantum state $\kett{\psi(\boldsymbol{\theta})}=U(\boldsymbol{\theta})\kett{\boldsymbol{0}}$. The probability distribution of bitstrings, and the energies corresponding to them, sampled after measurement of the state $\kett{\psi(\boldsymbol{\theta})}$ is used to compute a suitable objective. The goal is to identify the optimal parameters $\boldsymbol{\theta}$ that minimize the objective function, typically achieved through a classical optimization scheme. We use here a CVaR-based objective that is defined as the average of lower $\alpha$-tail of the energy distribution of the sampled bitstring. More formally, if the set of bitstring energies sampled and sorted in increasing order is denoted by $\{E_i\}$ where $i \in [1..K]$, then the CVaR($\alpha$) is defined as 
\begin{align}
    \frac{1}{\ceil[]{\alpha K}}\sum_{i=1}^{\ceil[]{\alpha K}} E_i.
\end{align}

When $\alpha=1$, CVaR($\alpha$) corresponds to the expectation value $\expec{\psi(\boldsymbol{\theta})}{H_p}{\psi(\boldsymbol{\theta})}$ of the problem Hamiltonian $H_p$, whereas when $\alpha \to 0$, only the minimum among sampled energies is considered. Employing an $\alpha$ value in the range $(0, 1]$ is expected to improve the convergence behavior of the classical optimizer~\cite{barkoutsos2020improving, kolotouros2105evolving}. More recently, authors in Ref.~\cite{barron2023provable} indicated CVaR to be robust against quantum hardware noise and provided bounds on the noise-free expectation values, placed by CVaR based on the noisy samples.

\subsection{Classical Optimizer}
Recently, several studies~\cite{lavrijsen2020classical,pellow2023qaoa} have examined the performance of classical optimizers for variational schemes, with an emphasis on robustness of such optimizers on the noisy near-term quantum hardware. In this work, we employ the NFT algorithm~\cite{nakanishi2020sequential} as the classical optimizer which was found to be relatively better under hardware noise~\cite{oliv2022evaluating, singh2023benchmarking}. NFT is a gradient-free classical optimizer that falls under the class of sequential minimal optimization algorithms such as Rotosolve~\cite{ostaszewski2021structure, li2024efficient, vidal2018calculus, parrish2019jacobi, wierichs2022general}. Under this scheme, each parameter $\theta_i$ is optimized sequentially. NFT uses the fact that for parameterized circuits with parameters sitting only on the single-qubit Pauli rotations (and fixed two-qubit gates), the expectation of an observable $\langle O\rangle$ can be expressed as an univariate sine function of $\theta_i$  with three unknown coefficients. Under the scheme, this sine function is constructed by evaluating $\langle O\rangle$ at three different values of $\theta_i$, while keeping rest of the parameters fixed. The optimal $\theta_i$ corresponds to the angle that minimizes the sine function. For convergence guarantees, unbiased estimate of $\langle O\rangle$ is needed. However, numerical simulations conducted in Ref.~\cite{nakanishi2020sequential} found NFT robust under finite sampling as well. In this work, we use CVaR values as a proxy to the expectation values. Based on the MPS-simulator VQE runs, we notice, NFT usually achieving convergence faster when CVaR with $\alpha < 1$ is used as a proxy to the expectation ($\alpha = 1$).          
\subsection{Ansatz}

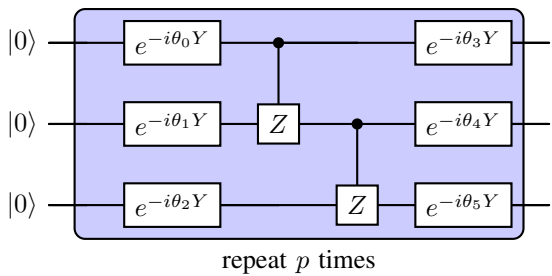
\begin{figure}[!ht]
\centering
\begin{quantikz}
\lstick{$\kett{0}$} & \qw \gategroup[wires=3, steps=5, style={solid, rounded corners, fill=blue!20, inner sep=1pt}, background, label style={label position=below, yshift=-0.5cm}]{repeat $p$ times} & \gate{e^{-i \theta_0 Y}} & \ctrl{1} & \qw & \gate{e^{-i \theta_3 Y}}  & \qw \\
\lstick{$\kett{0}$} & \qw & \gate{e^{-i \theta_1 Y}} & \gate{Z} & \ctrl{1} & \gate{e^{-i \theta_4 Y}}  &\qw \\
\lstick{$\kett{0}$} & \qw & \gate{e^{-i \theta_2 Y}} & \qw & \gate{Z} & \gate{e^{-i \theta_5 Y}} & \qw
\end{quantikz}
\caption{Two-local ansatz for a three-qubit case.}
\label{fig:ansatz}
\end{figure}

We use here the ``two-local'' ansatz involving single-qubit Pauli-Y rotations ($e^{-i\phi Y}$) and two-qubit control-Z gates, as depicted in  Fig.~\ref{fig:ansatz}. This configuration is compatible with the NFT optimizer.  For simulations, two-qubit gates are applied in a ``pairwise'' fashion composed of two layers. In the first layer, qubit $i$ is entangled with qubit $(i+1)$, for all even values of $i$ and in the second layer qubit $i$ is entangled with qubit $(i+1)$ for all odd values of $i$. For the hardware runs, the two-qubit gates are applied between qubits belonging to the actual hardware sub-graphs obtained after the hardware characterization steps described in Section~\ref{sec:selection}. 

\section{Results}\label{sec:results}
\subsection{Simulation runs}

\begin{figure}
\centering
\includegraphics[width=0.5\textwidth]{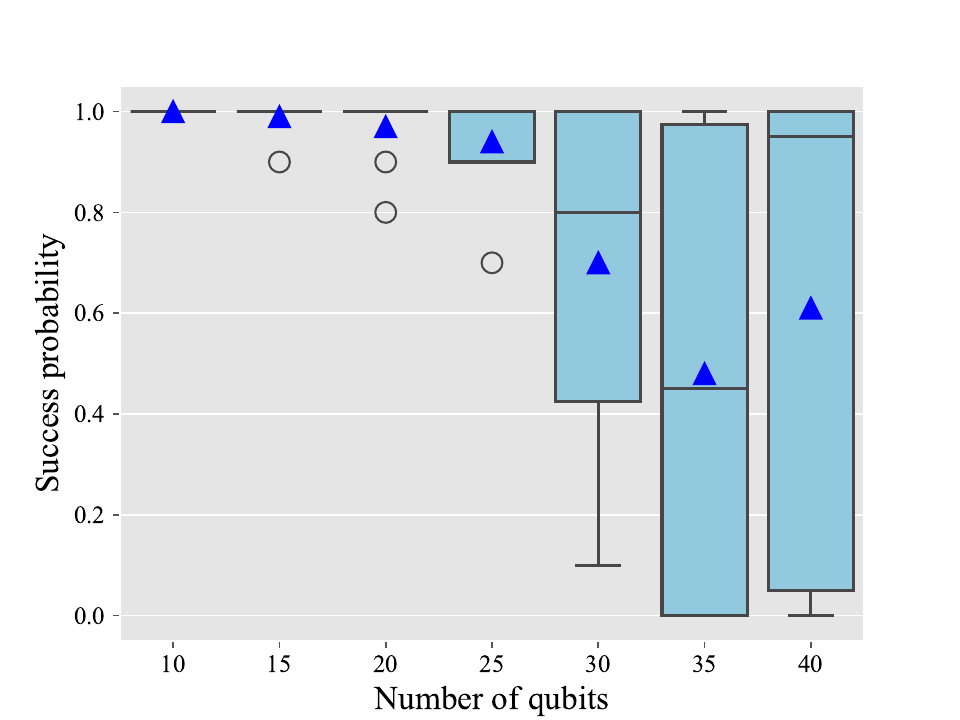}
\caption{Box plot depicting the success percentages for problem sizes between 10 and 40 qubits for CVaR-VQE runs on the MPS simulator. For each problem size, simulations were conducted on ten distinct sequences. For each sequence, ten independent CVaR-VQE trials were performed. The success percentage is calculated as the number of successful runs over ten trials. The average over different sequences is indicated by blue triangles.}
\label{fig:sim prob}
\end{figure}

\begin{figure}
\centering
\includegraphics[width=0.5\textwidth]{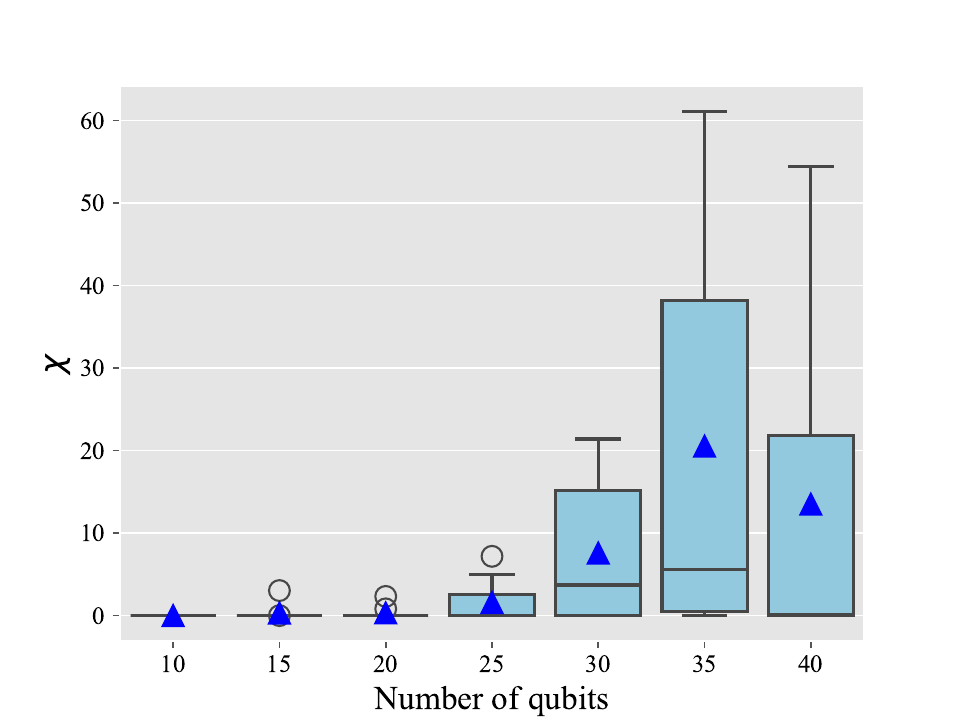}
\caption{Box plot of the optimality gap $\chi$ for problem sizes between 10 and 40 qubits for CVaR-VQE runs on the MPS simulator. The averaging scheme is the same as described in Fig.~\ref{fig:sim prob}.}
\label{fig:sim chi}
\end{figure}

We perform noise-free simulation of CVaR-VQE using the matrix-product state (MPS)~\cite{vidal2003efficient, schollwock2011density} simulator provided in Qiskit~\cite{Qiskit}. Since the computational cost of MPS-based simulator grows exponentially with entanglement~\cite{vidal2003efficient}, we limit our simulations to shallow-depth circuits. We use two layers ($p=2$) of the ``two-local'' ansatz as shown in Fig.~\ref{fig:ansatz}. After examining several different values of the CVaR parameter $\alpha$ (see Section \ref{cvar-vqe}) on relatively small problem sizes, we set it equal to 0.1 for all the simulation runs. The number of shots $N_{\mathrm{shots}}$ used in the simulations was $2^5$ and $2^{10}$ for the 10- and 15-qubit problems, respectively, and $2^{13}$ for the rest of the problems. The NFT optimizer evaluates $n_{\mathrm{eval}}=3$ sets of circuit parameters at each iteration. Each iteration consists of $n_{\mathrm{eval}} \times N_{\mathrm{shots}}$ circuit calls. The optimization scheme is executed for $N_{\mathrm{iter}}$ iterations until sufficient convergence of the CVaR values is observed, typically in the range of 100 to 200 iterations.

Because classical optimizers like NFT are prone to getting stuck in local minima, we run $N_{\mathrm{trial}} = 10$ independent trials of CVaR-VQE for each problem instance. Each trial run starts with a set of circuit parameters initialized randomly from a uniform distribution. We examine 10 different mRNA sequences for each qubit size.

Fig.~\ref{fig:sim prob} presents a box plot illustrating the success probability across various problem instances ranging from 10 and 40 qubits. A run is deemed successful if the bitstring with the lowest energy, identical to the solution obtained by CPLEX, is found within the samples collected after measuring the final quantum state $\kett{\psi(\boldsymbol{\theta_f})}$, where $\boldsymbol{\theta_f}$ denotes the circuit parameters obtained after $N_{\mathrm{iter}}$ iterations. The success probability $p_{\mathrm{succ}}$ is then defined as $N_{\mathrm{succ}}/N_{\mathrm{trial}}$, where $N_{\mathrm{succ}}$ is the number of successful trials. As evident from Fig.~\ref{fig:sim prob}, the CVaR-VQE algorithm demonstrates robust performance, with the average success probability in the range from 0.4 to 1.0. 

In addition to the success probability, we also use  the optimality gap as a metric to assess the quality of the solutions obtained from CVaR-VQE. The optimality gap $\chi$ is defined as 

\begin{align}
    \chi = \frac{\abs{F(\boldsymbol{\theta}_{\mathrm{f}})_{\mathrm{low}}- F_0}}{\abs{F_0}} \times 100\,,
\end{align}
where $F(\boldsymbol{\theta}_{\mathrm{f}})_{\mathrm{low}}$ represents the lowest value of the objective function, as defined in Eq.~\ref{eq:qubo}, found among the samples at the end of $N_{\mathrm{iter}}$ iterations. $F_0$ denotes the value of the objective function of the optimal solution found by CPLEX. $\chi$ is zero when the lowest energy found by the quantum optimization matches the one found by CPLEX, and greater than zero otherwise. The smaller the deviation from zero, the better the quality of solutions obtained using CVaR-VQE.

Fig.~\ref{fig:sim chi} shows a box plot of the optimality gap $\chi_{\mathrm{avg}}$, averaged over 10 independent trials for each sequence, for qubits sizes ranging from 10 to 40. The CVaR-VQE algorithm demonstrates high performance, achieving an average optimality gap of less than 20$\%$.

Our results indicate that the performance of the algorithm deteriorates as the problem size increases. For the constant-depth circuits employed in this study, such a decline with increasing problem size is anticipated. We hypothesize that the performance of CVaR-VQE can be further enhanced by increasing the number of layers in the ansatz, optimizing the choice of $\alpha$, or by incorporating other recent advancements in CVaR~\cite{kolotouros2022evolving}. We intend to investigate these aspects of the algorithm in future research.

\subsection{Hardware runs}\label{sec:hardware}

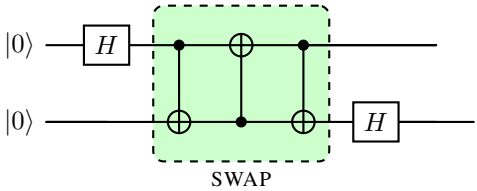
\begin{figure}[!ht]
\centering
\begin{quantikz}
\lstick{$\kett{0}$} & \gate{H} &
    \ctrl{1}\gategroup[2,steps=3,style={dashed,rounded
    corners,fill=green!20, inner
    xsep=2pt},background,label style={label
    position=below,anchor=north,yshift=-0.2cm}]{{\sc
    swap}} & \targ{} & \ctrl{1} & \qw & \qw \\
\lstick{$\kett{0}$} & \qw & \targ{} & \ctrl{-1} & \targ{} & \gate{H} & \qw & 
\end{quantikz}
\caption{Quantum circuit for noise characterization.}
\label{fig:characterization}
\end{figure}

In this Section, we provide details about the hardware runs carried out on the IBM Eagle and Heron processors, including \texttt{ibm\_brisbane}, \texttt{ibm\_osaka}, and \texttt{ibm\_torino}.

\subsubsection{Qubit Selection}
\label{sec:selection}

\begin{figure*}
  \begin{subfigure}[t]{\textwidth}
    \includegraphics[width=\textwidth]{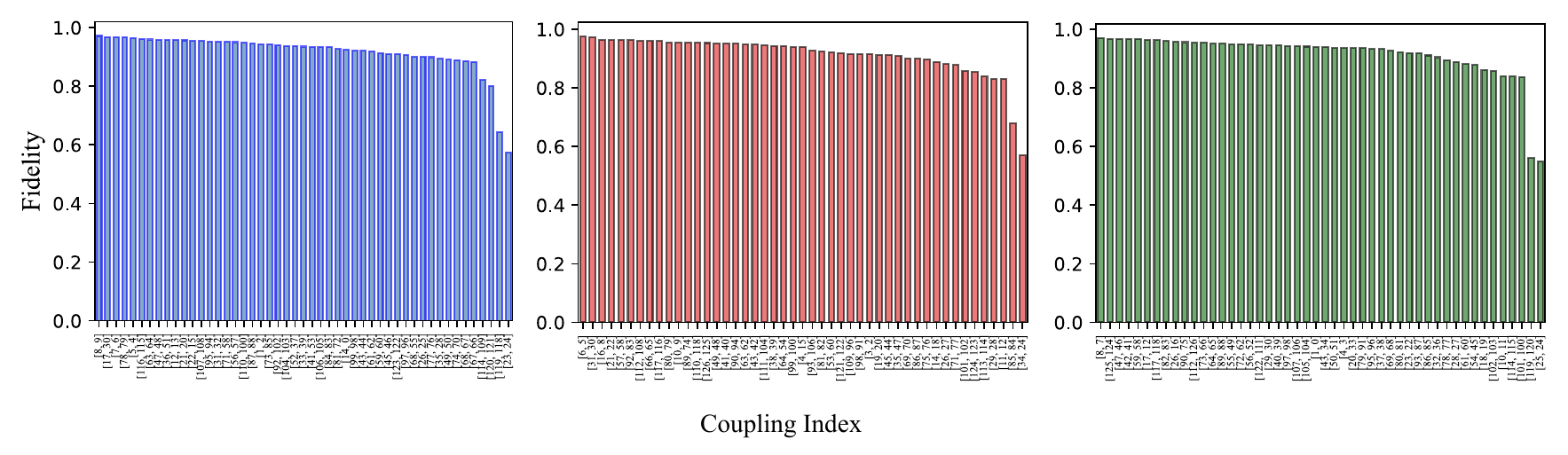}
    \caption{}
    \label{fig:fidelities}
  \end{subfigure}\hfill
  \begin{subfigure}[t]{\textwidth}
  \centering
    \includegraphics[width=0.5\textwidth, height=180pt]{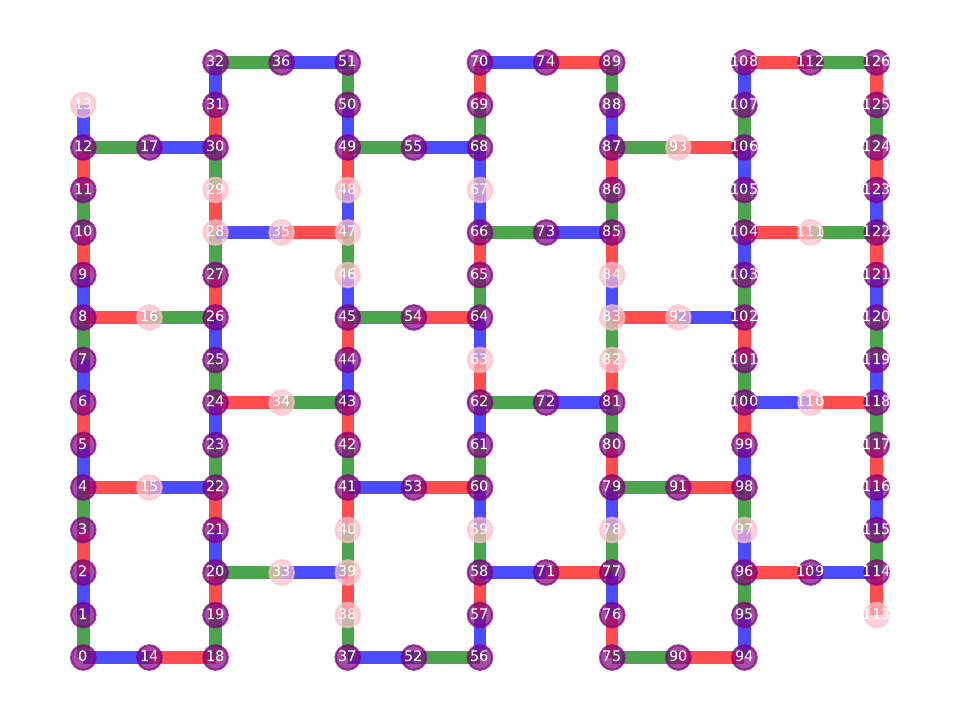}
    \caption{}
    \label{fig:layout}
  \end{subfigure}
  \caption{Panel a) shows the fidelity of different couplings across the three layers obtained during the hardware characterization scheme detailed in Section~\ref{sec:selection}. Panel b) shows the heavy-hex layout on \texttt{ibm\_brisbane} and \texttt{ibm\_osaka} with nodes on the graph representing qubits and edges representing qubit couplings. The couplings belonging to the three layers are color coded as blue, red, and, green, matching the fidelity plots in a). The purple colored nodes represent the 100 qubit chain considered for selection of qubits on \texttt{ibm\_brisbane} based on the fidelity values.} 
  \label{fig:benchmark}
\end{figure*}

Due to variations in the noise characteristics across the physical chip of a quantum computer, mapping problem qubits to the hardware sub-graphs with smaller error rates becomes crucial.  To account for the cross-talk, similar in spirit to the layered fidelity benchmarking scheme~\cite{mckay2023benchmarking}, we examine fidelity across circuit layers containing disjoint set of qubit couplings. We implement a simple two-qubit circuit (see Fig.~\ref{fig:characterization}) across couplings on each layer separately. Note, that the three layers with couplings of color blue, red, and green in Fig.~\ref{fig:benchmark} cover all of the neighboring interactions present on the quantum device. Fidelity between ideal ($00$ bitstring) and the distribution obtained from the quantum computer is indicated in Fig.~\ref{fig:benchmark} for the three layers. Next, in order to select the set of ``good'' qubits, we consider a 1D-chain of 100 qubits, color coded as nodes in magenta in Fig.~\ref{fig:benchmark}~\cite{mckay2023benchmarking}. Across the chain, we exclude qubits connected to couplings with fidelities not meeting a set threshold. For our runs we set this fidelity threshold to 0.85. If at the end of the process, we still need more qubits, we extend our search to the rest of the couplings not covered by the 100-qubit chain. We employ the described process of qubit selection on all the devices once before starting the CVaR-VQE runs.  


\begin{table}[!ht]
\begin{center}
\caption{Quantum device circuit and run attributes. Max iterations column indicates the maximum number of classical iteration steps carried out by the NFT optimizer. The two-qubit Echoed Cross Resonant (ECR) gate is native to \texttt{ibm\_brisbane} and \texttt{ibm\_osaka}. The Control-Z (CZ) gate is native to \texttt{ibm\_torino}.}
    
\label{circuit attributes}
\begin{tabular}{|r||*{3}{p{9mm}}|c|}
\hline
& \multicolumn{3}{|c|}{\texttt{ibm\_brisbane}/\texttt{ibm\_osaka}}
& \texttt{ibm\_torino}  \\
\hline
\hline
Qubits & 26 & 40 & 50 & 80\\
Circuit depth & 18 &20& 20& 17 \\
Max iterations & 104 & 320 & 600 & --\\
ECR$\slash$CZ Count & 25 & 39 & 49 & 158 \\
\hline
\end{tabular}
\end{center}
\end{table}

\begin{figure}
\begin{subfigure}{.5\linewidth}
\centering
\includegraphics[width=1\textwidth]{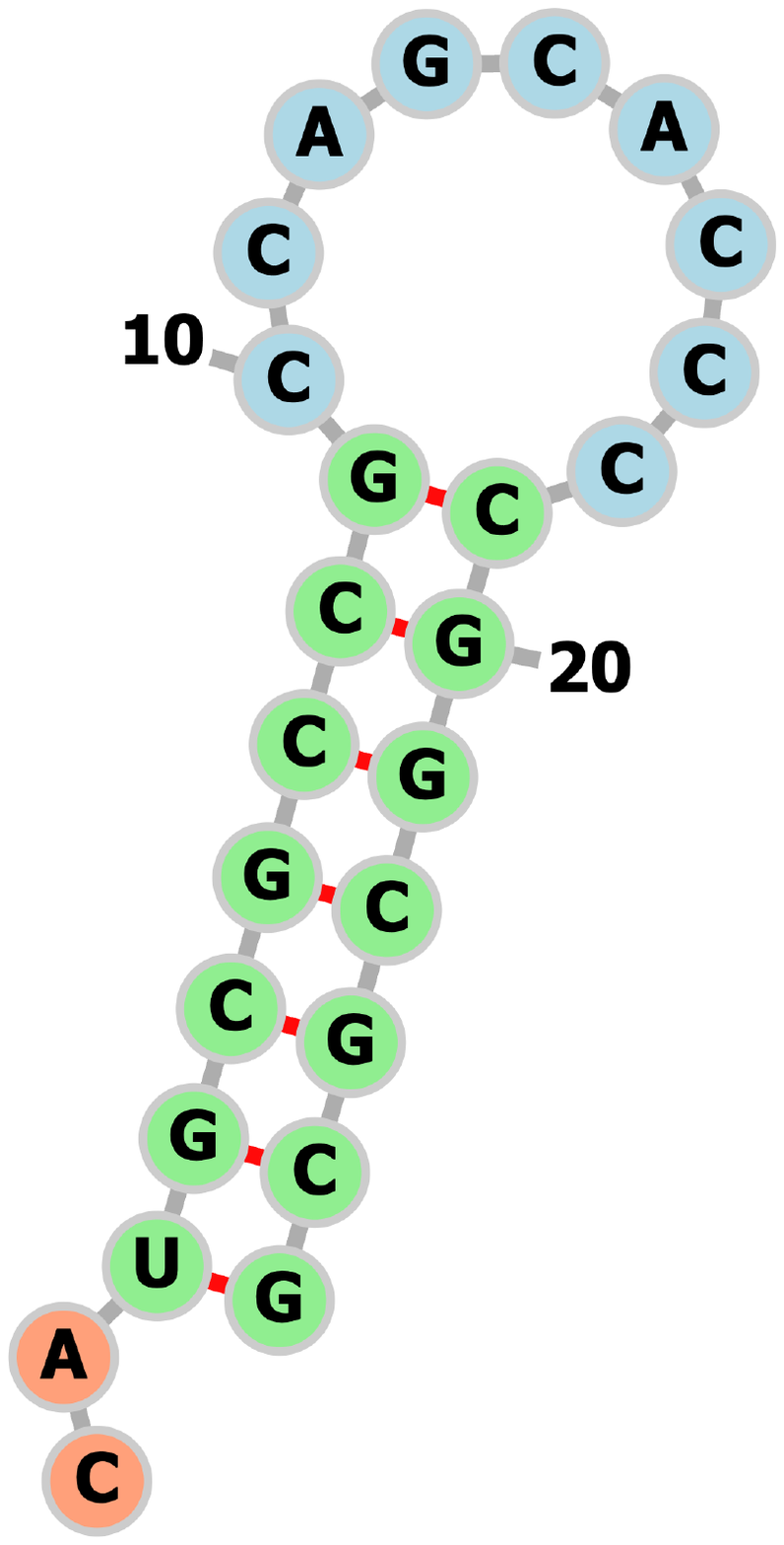}
\caption{26 qubits, 25 nucleotides}
\label{fig:40 qubits}
\end{subfigure}%
\begin{subfigure}{.5\linewidth}
\centering
\includegraphics[width=1.2\textwidth]{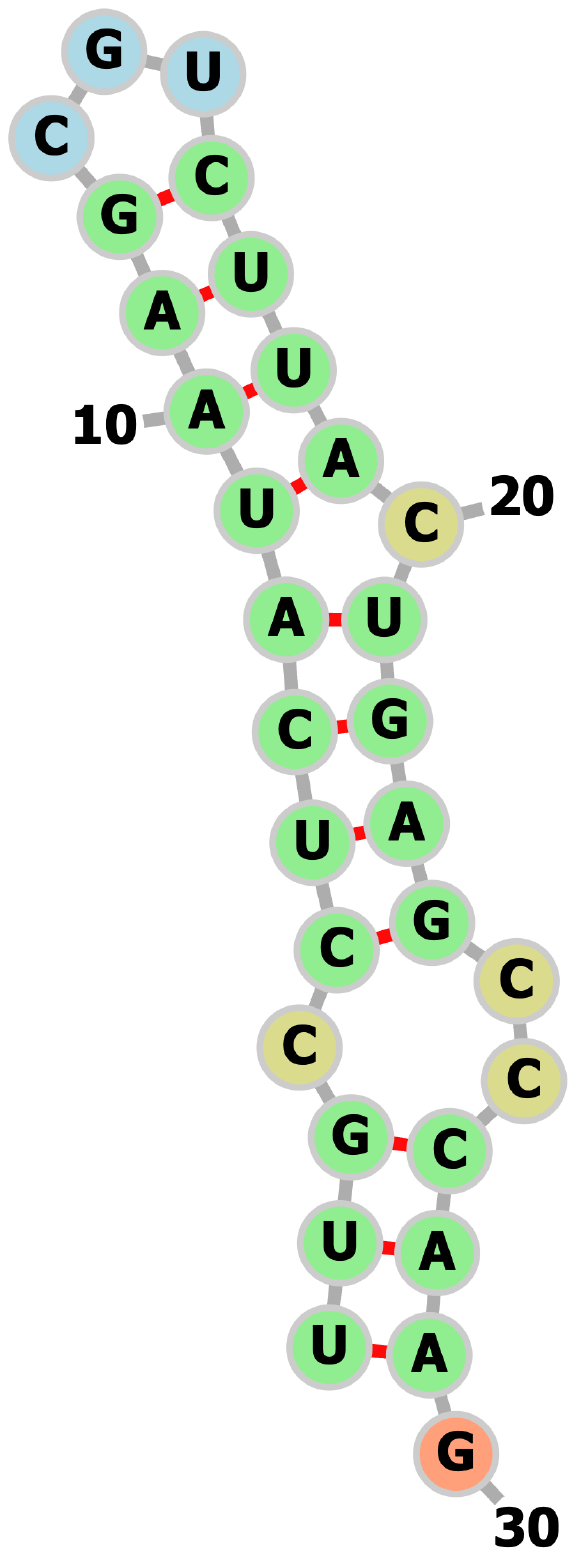}
\caption{40 qubits, 30 nucleotides}
\label{fig:50 qubits, xx sequences}
\end{subfigure}\\[1ex]
\begin{subfigure}{1.0\linewidth}
\centering
\includegraphics[width=0.6\textwidth]{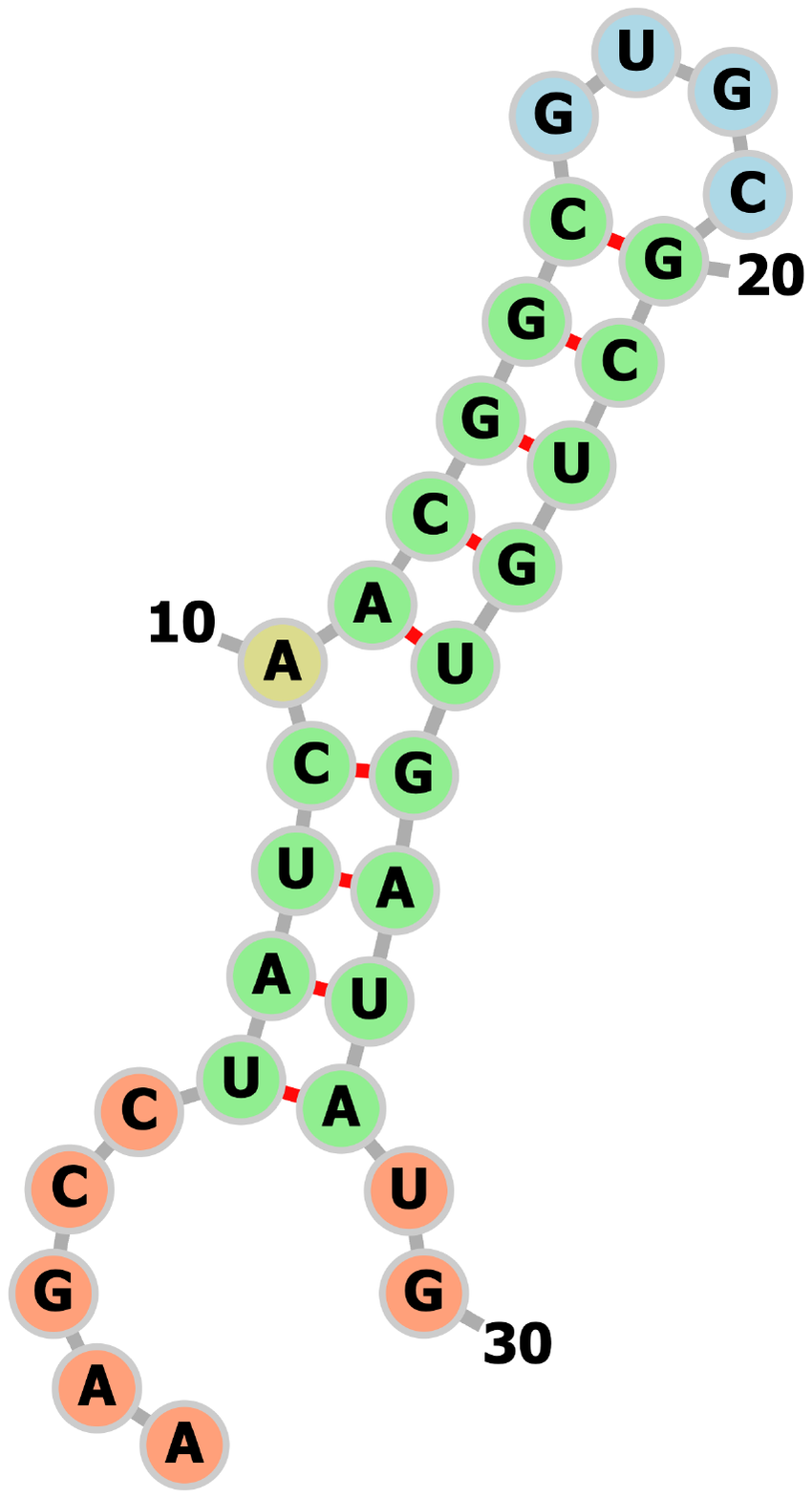}
\caption{50 qubits, 30 nucleotides}
\label{fig:40 qubits2}
\end{subfigure}%
\caption{Figures show actual optimal sequences based on the lowest energy bitstring found during hardware runs.}
\label{fig:optimal sequences}
\end{figure}

\subsubsection{Error suppression and error mitigation}
We use Dynamical Decoupling (DD)~\cite{viola1998dynamical, zanardi1999symmetrizing, vitali1999using, ezzell2023dynamical}, which adds pulses during the idle times of the circuit and helps to suppress errors without any additional overhead. Additionally, we use a version of Matrix-free Measurement Mitigation (M3)~\cite{nation2021scalable} for readout error mitigation. Instead of working based off a full assignment matrix of size $2^n$, M3 uses a much smaller subspace formed by the noisy bitstrings, allowing it to usually handle large system efficiently~\cite{nation2021scalable}. In order to keep the error mitigation overhead to ``minimal'', we do not consider mitigation schemes like ZNE~\cite{kim2023evidence, kandala2019error, li2017efficient} and PEC~\cite{van2023probabilistic} in this work. Interestingly, as pointed out in Ref.~\cite{barron2023provable}, CVaR-based schemes with a relatively small sampling overhead  may still be able to track the noise-free expectation values~\cite{barron2023provable}. We plan to explore these aspects of the CVaR algorithm in near future. Additionally, we set \textrm{optimization\_level=3} under the Qiskit's Sampler V1 primitive runtime options to optimize circuits, which gives rise to transpiled circuits depths in the range of 17-20 (see Table~\ref{circuit attributes}).

\begin{figure}[!ht]
\centering
\includegraphics[width=.5\textwidth]{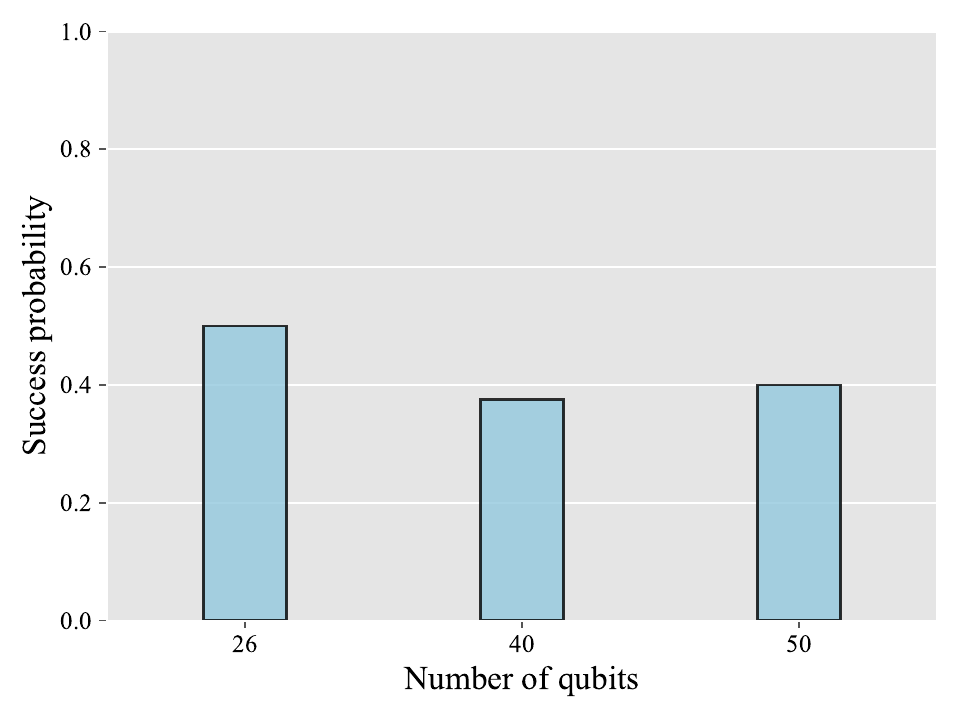}
\caption{Success probability for different number of qubits for the hardware runs.}
\label{fig:hardware prob}
\end{figure}

\begin{figure}[!ht]
\centering
\includegraphics[width=.55\textwidth]{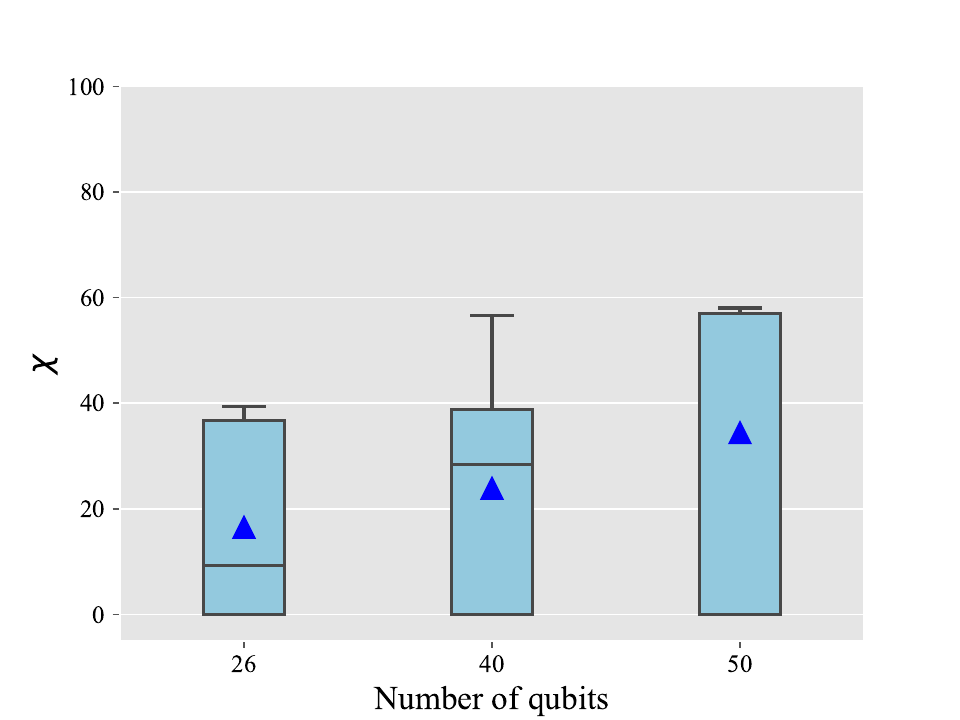}
\caption{Box plot of the optimality gap $\chi$ for problem sizes in the range of 10 to 40 qubits executed on quantum hardware. Average $\chi$ over $N_{\mathrm{trial}}$ independent runs is shown using blue triangles.}
\label{fig:hardware chi}
\end{figure}

\subsubsection{Hardware Results}
On the hardware, we study 3 mRNA sequences: (a) of length 25, represented by the problem instance with 26 qubits, (b) of length 30, with 40 qubits, and (c) of length 30, with 50 qubits. We run 8, 8, and 5 independent CVaR-VQE trials for these instances, correspondingly, using $\alpha=0.25$ and $p=1$.

The success probability of CVaR-VQE experiments on quantum hardware is shown in Fig.~\ref{fig:hardware prob}. It should be noted that hardware experiments were conducted for only one sequence of each problem size; therefore, the success probability is computed as a cumulative quantity over 10 trials. As indicated by the plot, the success probability across different qubit sizes is above $\sim0.38$.  The average optimality gap $\chi_{\mathrm{avg}}$ of CVaR-VQE experiments on the hardware is shown in Fig.~\ref{fig:hardware chi} and is below $\sim$34$\%$. We set the number of shots for each circuit to $2^8$ for the 26-qubit problem and $2^{13}$ for the larger problem instances. The maximum number of iterations required to find the optimal parameters on the hardware is reported in the Table~\ref{circuit attributes}. Some trials took fewer than the maximum number of iterations reported. The lowest energy sequences found are shown in Fig.~\ref{fig:optimal sequences}. 

\begin{figure}
\centering
\begin{subfigure}{1.0\linewidth}
\includegraphics[width=\textwidth]{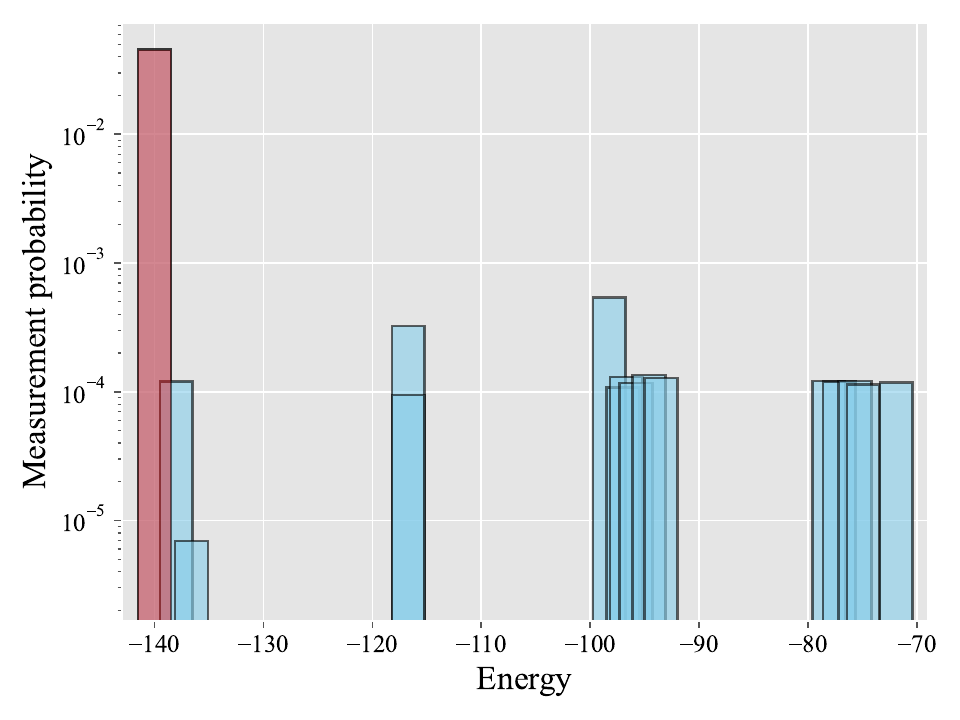}
\end{subfigure}%

\begin{subfigure}{1.0\linewidth}
\centering
\includegraphics[width=0.8\textwidth]{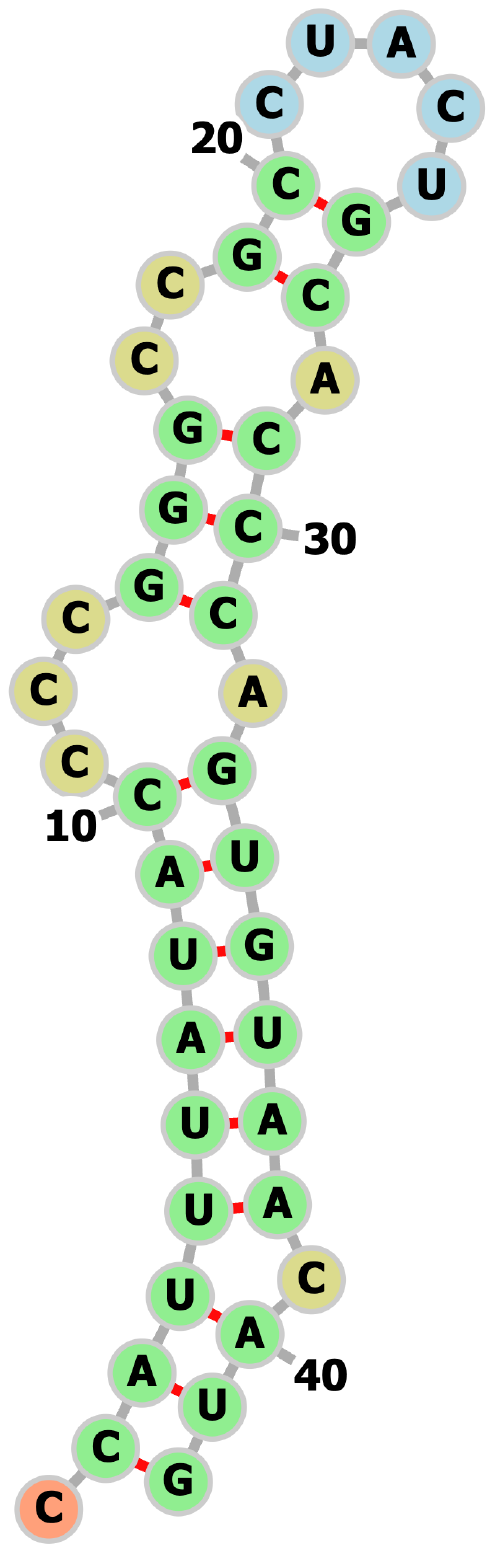}
\end{subfigure}
\caption{(a) Measurement probability of sampled bitstrings plotted against bitstring energies for the 80-qubit problem. Red bar indicates the probability corresponding to the lowest energy bitstring. (b) Optimal folded structure of the 42-nucleotide, 80-qubit mRNA sequence based on the corresponding lowest energy bitstring found by the hardware run.}
\label{fig:80 qubit}
\end{figure}

For the 42-nucleotide, 80-qubit mRNA problem instance, we do not carry out circuit parameter optimization on the hardware. The optimal parameters for the circuit were obtained by running CVaR-VQE using the MPS simulator with the $\alpha=0.1$ and $p=2$. We use the optimized parameters to run circuits on the recently introduced \texttt{ibm\_torino} device featuring the 133-qubit tunable-coupler Heron processor~\cite{ibm2023newsroom}. The resulting circuit has depth 17 and contains 158 two-qubit CZ gates. We plot the distribution of measurement probabilities of the sampled bitstrings with the corresponding energies in Fig.~\ref{fig:80 qubit}. The plot illustrates that the lowest energy bitstring with energy value of -140, which matches the lowest energy bitstring found by CPLEX, is sampled with a probability separated from the rest of the sampling probabilities by more than one order of magnitude. The measurement probability of the lowest bitstring is $0.045$, hence, we need at most $1/0.045 \sim 20$ samples to find the lowest energy bitstring. The optimal sequence found is shown in Fig.~\ref{fig:80 qubit}. Note, the offline training of circuit parameters using MPS simulator was only possible because of the classical simulatability of such circuits. As pointed out in Refs.~\cite{sack2024large, patra2024efficient, tindall2024efficient, liao2023simulation, anand2023classical, beguvsic2023fast, beguvsic2024fast, rudolph2023classical, shao2023simulating}, the heavy-hex lattice-based circuits characterized by depths we employ in this study can be efficiently simulated using classical schemes. Our results point to an exciting regime where relatively large scale optimization-based problems, approaching utility-scale, can also be successfully executed on the noisy quantum hardware and can produce reliable results using low-overhead error mitigation schemes.       

\subsection{Feasibility of hardware runs}
\label{feasibility}

Unlike schemes targeting fault-tolerant quantum devices, where quantum resource scaling arguments are relatively straightforward to derive, establishing such arguments for variational schemes is more challenging~\cite{scriva2024challenges}. If $p$ layers of an ansatz are employed, and each circuit layer requires time $T$ to execute, the total time taken is approximately $N_{\mathrm{circuit}} \times p \times T$, where $N_{\mathrm{circuit}}$ represents the total number of circuits executed. Given that this study utilizes fixed-depth circuits and assuming the execution time $T$ remains relatively invariant with problem size, the primary factor that necessitates closer examination is $N_{\mathrm{circuit}}$. 

The number of circuits $N_{\mathrm{circuit}}$ executed during the scheme is determined by the expression $1/p_{\mathrm{succ}} \times N_{\mathrm{iter}} \times N_{\mathrm{shots}} \times n_{\mathrm{eval}}$, where $n_{\mathrm{eval}}$ represents the number of different sets of circuit parameters evaluated at each iteration to suggest a new set of circuit parameters $\boldsymbol{\theta}$\footnote{For the NFT optimizer, $n_{\mathrm{eval}} = 3$}. To assess the feasibility of solving structure prediction problems on a quantum computer, it is crucial that none of the three factors $\{1/p_{\mathrm{succ}}, N_{\mathrm{iter}}, N_{\mathrm{shots}}\}$, with $n_{\mathrm{eval}}$ assumed constant, scale exponentially with the number of qubits. A concerning trend observed is the decline in success probabilities $p_{\mathrm{succ}}$ as the problem size increases. However, based on the results, it is difficult to definitively conclude that the decline is exponential. Additionally, since the choice of algorithm dictates the lower bound on $p_{\mathrm{succ}}$, there is potential for improvement through algorithmic advances. Unfortunately, beyond a certain regime of classical simulatability, performance comparison of different optimization schemes can only be carried out in conjunction with the quantum hardware.

\section{Conclusions and outlook}\label{sec:conclusions}

In this work, we examined the feasibility of solving secondary structure prediction-based optimization problems of RNA sequences on a quantum computer. Our results demonstrate that optimization schemes like CVaR-VQE running on noisy quantum hardware with careful calibration and nominal error-mitigation, can produce sufficiently reliable outputs even for relatively large problem instances. However, the scalability of these methods to larger problem sizes exceeding 100 qubits remains an open question that warrants further investigation. As quantum hardware continues to advance, any improvements in algorithmic techniques will further bolster prospects of using quantum computers for optimization. This study, which primarily focused on the capabilities of utility-scale quantum hardware, only considered a subset of available quantum approaches; however, our future work will explore more advanced optimization techniques.

Our results indicate that quantum computing offers valuable insights into RNA folding mechanisms and enables rapid identification of functional RNA structures. This work lays the groundwork for future research in quantum-assisted bioinformatics and highlights the importance of developing scalable quantum algorithms for real-world biological problems.  

\section*{Acknowledgments}

The authors would like to thank Yashrajsinh Jadeja and Haining Lin for their valuable assistance in acquiring test sequences and for their contributions to the prediction of classical RNA secondary structures.

\bibliographystyle{myIEEEtran}
\bibliography{ref}

\end{document}